\documentclass[10pt,conference]{IEEEtran}
\pdfoutput=1
\usepackage[utf8]{inputenc}
\usepackage[T1]{fontenc}
\usepackage{graphicx}
\usepackage[usenames,dvipsnames,svgnames,table]{xcolor}
\usepackage[hidelinks]{hyperref}
\usepackage{amssymb}
\usepackage{pifont}
\usepackage{acronym}
\usepackage{tikz}
\usepackage[frozencache=true,newfloat=true]{minted}
\usepackage[binary-units,per-mode=symbol]{siunitx}
\usepackage[backend=bibtex,sorting=none,style=ieee,isbn=false]{biblatex}
\usepackage[abbreviations]{foreign}

\SetupFloatingEnvironment{listing}{placement=tp}
\setminted{fontsize=\small, framesep=7pt, frame=topline, framerule=0pt, style=bw}

\bibliography{biblio}
\renewcommand*{\bibfont}{\fontsize{9}{5}}

\def\figurewidth{230pt}

\hypersetup{
	pdftitle={SGX-Aware Container Orchestration for Heterogeneous Clusters},
	pdfauthor={Sébastien Vaucher, Rafael Pires, Pascal Felber, Marcelo Pasin, Valerio Schiavoni, Christof Fetzer},
}

\newcommand\copyrighttext{\footnotesize \textcopyright{}~2018 IEEE.
	Personal use of this material is permitted.
	Permission from IEEE must be obtained for all other uses,
	in any current or future media, including reprinting/republishing this
	material for advertising or promotional purposes, creating new collective
	works, for resale or redistribution to servers or
	lists, or reuse of any copyrighted component of this work in other works.
	Presented in the \href{http://icdcs2018.ocg.at}{38th IEEE International Conference on Distributed Computing Systems (ICDCS '18)}. The final version of this paper is available under \textsc{doi}: \href{https://doi.org/10.1109/ICDCS.2018.00076}{10.1109/ICDCS.2018.00076}.}

\acrodef{SGX}{Software Guard Extensions}
\acrodef{TEE}{Trusted Execution Environment}
\acrodef{EPC}{Enclave Page Cache}
\acrodef{CPU}{Central Processing Unit}
\acrodef{MEE}{Memory Encryption Engine}
\acrodef{LE}{Launch Enclave}
\acrodef{QE}{Quoting Enclave}
\acrodef{PE}{Provisioning Enclave}
\acrodef{AESM}{Application Enclave Service Manager}
\acrodef{API}{Application Programming Interface}
\acrodef{gRPC}{Google Remote Procedure Calls}
\acrodef{GPU}{Graphics Processing Unit}
\acrodef{SDK}{Software Development Kit}
\acrodef{PSW}{Platform Software}
\acrodef{FPGA}{Field-Programmable Gate Array}
\acrodef{CDF}{Cumulative Distribution Function}
\acrodef{TCB}{Trusted Computing Base}
\acrodef{SMM}{System Management Mode}
\acrodef{PRM}{Processor Reserved Memory}
\acrodef{UEFI}{Unified Extensible Firmware Interface}
\acrodef{FCFS}{first-come first-served}
\acrodef{IaaS}{Infrastructure as a Service}
\acrodef{SEV}{Secure Encrypted Virtualization}
\acrodef{TPM}{Trusted Platform Module}
 
\renewbibmacro{finentry}{
	\iffieldequalstr{entrykey}{Brenner:2016:SecureKeeper}
	{\finentry\newpage}
	{\finentry}}

\begin{document}

\title{SGX-Aware Container Orchestration\\for Heterogeneous Clusters}
\date{}

\author{\IEEEauthorblockN{Sébastien Vaucher\IEEEauthorrefmark{1}, Rafael Pires\IEEEauthorrefmark{1}, Pascal Felber\IEEEauthorrefmark{1}, Marcelo Pasin\IEEEauthorrefmark{1}, Valerio Schiavoni\IEEEauthorrefmark{1} and Christof Fetzer\IEEEauthorrefmark{2}}
\IEEEauthorblockA{\IEEEauthorrefmark{1}University of Neuchâtel, Switzerland. E-mail: \url{first.last@unine.ch}}
\IEEEauthorblockA{\IEEEauthorrefmark{2}Technical University of Dresden, Germany. E-mail: \url{christof.fetzer@tu-dresden.de}}
}

\maketitle

\begin{abstract}
Containers are becoming the \emph{de facto} standard to package and deploy applications and micro-services in the cloud.
Several cloud providers (\eg Amazon, Google, Microsoft) begin to offer native support on their infrastructure by integrating container orchestration tools within their cloud offering.
At the same time, the security guarantees that containers offer to applications remain questionable.
Customers still need to trust their cloud provider with respect to data and code integrity.
The recent introduction by Intel of \ac{SGX} into the mass market offers an alternative to developers, who can now execute their code in a hardware-secured environment without trusting the cloud provider.

This paper provides insights regarding the support of \ac{SGX} inside Kubernetes, an industry-standard container orchestrator.
We present our contributions across the whole stack supporting execution of SGX-enabled containers.
We provide details regarding the architecture of the scheduler and its monitoring framework, the underlying operating system support and the required kernel driver extensions.
We evaluate our complete implementation on a private cluster using the real-world Google Borg traces.
Our experiments highlight the performance trade-offs that will be encountered when deploying \ac{SGX}-enabled micro-services in the cloud.
\end{abstract}

\begin{tikzpicture}[remember picture,overlay]
\node[anchor=south,yshift=10pt] at (current page.south) {\fbox{\parbox{\dimexpr\textwidth-\fboxsep-\fboxrule\relax}{\copyrighttext}}};
\end{tikzpicture}%
\acresetall

\section{Introduction}
\label{sec:introduction}

There has been a steady trend over the last few years~\cite{docker-trend} for applications to be packaged and deployed in the form of containers, such as Docker~\cite{anderson2015docker,merkel2014docker}.
Containers offer reproducible execution environments, light lifecycle management, and closer-to-metal performance than classical virtual machine deployments.
Container registries---public ones such as Docker Hub~\cite{dockerhub}, or private ones such as Google Cloud Registry~\cite{google-registry}---hold and serve thousands of container images, ready to be used.
Hence, developers can simply rely on cloud infrastructures to deploy services of various nature.
As such, containers are used for all kind of services, from simple key-value stores (\eg Redis~\cite{redis-hub}), to web servers (Apache~\cite{redis-hub}); from  full-fledged relational databases (MySQL~\cite{mysql-hub}) to scalable discovery services~\cite{consul-hub}.
Major cloud operators natively support container deployments in their infrastructure (\eg Google Container Engine~\cite{googlecloudcomputecontainers} or Amazon EC2 Container Service~\cite{ec2container}).
Similarly, it is straightforward to set up a container cluster on private premises, leveraging popular~\cite{container-trends} container orchestrators (or schedulers) such as Kubernetes or Docker Swarm.

Without special care, containers are exposed to critical security threats.
For instance, the cloud infrastructure could be compromised by malicious actors or software bugs.
This is especially true when containers are deployed over public cloud infrastructures, but it also holds true for the case of a deployment on private premises that may have been corrupted by malicious actors or software.
Hence, service deployers are left with no other choice than to trust the infrastructure provider and the complete software stack (including the underlying operating system, kernel libraries, etc.).
Similarly, they must face the risk that a compromised component can lead to severe data leakage~\cite{gao:2017:containerleaks,takabi2010security}. 

To overcome these issues, one could adopt solutions based on homomorphic cryptosystems~\cite{gentry2009fully}. 
However, their performance is several orders of magnitude slower than native systems, and as of today, these solutions are still impractical for real-world deployment and adoption~\cite{gentry2011implementing}.

The introduction of Intel \ac{SGX}~\cite{costan2016intel} into the mass market with the Intel Skylake family of processors~\cite{rotem2015intel} offers a promising alternative.
The availability of SGX allows the deployers to distrust the cloud operator and rely instead on hardware protection mechanisms, hence drastically reducing the \ac{TCB}.
Moreover, programs execute at almost native speed~\cite{arnautov2016scone}.
Cloud providers are starting to offer SGX-enabled \ac{IaaS} solutions to end-users.
One example is Microsoft Azure Confidential~\cite{azureconfidential}.
We expect other cloud providers to introduce similar offerings in the short-term.

Deploying and orchestrating containers on a heterogeneous cluster, with a mix of machines with and without SGX capabilities, presents its own set of specific challenges.
The containers that require SGX will contend on the availability of dedicated memory (see \autoref{sec:background}).
Therefore, the monitoring infrastructure that feeds the scheduler master with resource usage metrics must keep track of the demanded SGX memory requests and schedule the containers accordingly. 
Unfortunately, none of the existing container orchestrators offer native support to provide runtime insights about the resources used by SGX containers.
Notably, all of them rely on statically-provided information given by the users upon deployment.
This information can be malformed or non-conforming to the real usage of the containers, and henceforth leading to over- or under-allocations.

In the proposed context, our contributions are the following.
We propose an SGX-aware architecture for orchestrating containers.
Fitting in this architecture, we offer an open-source vertical implementation~\cite{github:sgx-scheduler} of the required system support, including modifications to the Linux driver for \ac{SGX} as well as a Kubernetes \textit{device plugin}.
Further, we show that it will be possible to drastically reduce the waiting time of the submitted jobs by exploiting future versions of SGX, as they will offer better control over the size of the dedicated memory.
Finally, we demonstrate that our design and implementation are sound with a detailed evaluation using the Google Borg traces~\cite{clusterdata:Wilkes2011}.

The rest of the paper is organized as follows.
We provide a short introduction to Intel SGX in \autoref{sec:background}.
In \autoref{sec:model}, we describe our trust model and assumptions.
We describe the architecture in \autoref{sec:architecture} and our prototype implementation in \autoref{sec:implementation}, followed by its evaluation in \autoref{sec:evaluation}.
We survey the related work in \autoref{sec:related}, before concluding and describing how we plan to extend this work in \autoref{sec:conclusion}.
 
\section{Background on Intel SGX}
\label{sec:background}

The design of our system revolves around the availability of Intel \ac{SGX} on hosts.
It consists in \iac{TEE} recently introduced by the Skylake family of Intel processors, similar in spirit to ARM TrustZone~\cite{arm2009security}.
Applications create secure \emph{enclaves} to protect the integrity and confidentiality of the code being executed with associated data.

\begin{figure}[!t]
  \centering
  \includegraphics[width=\linewidth]{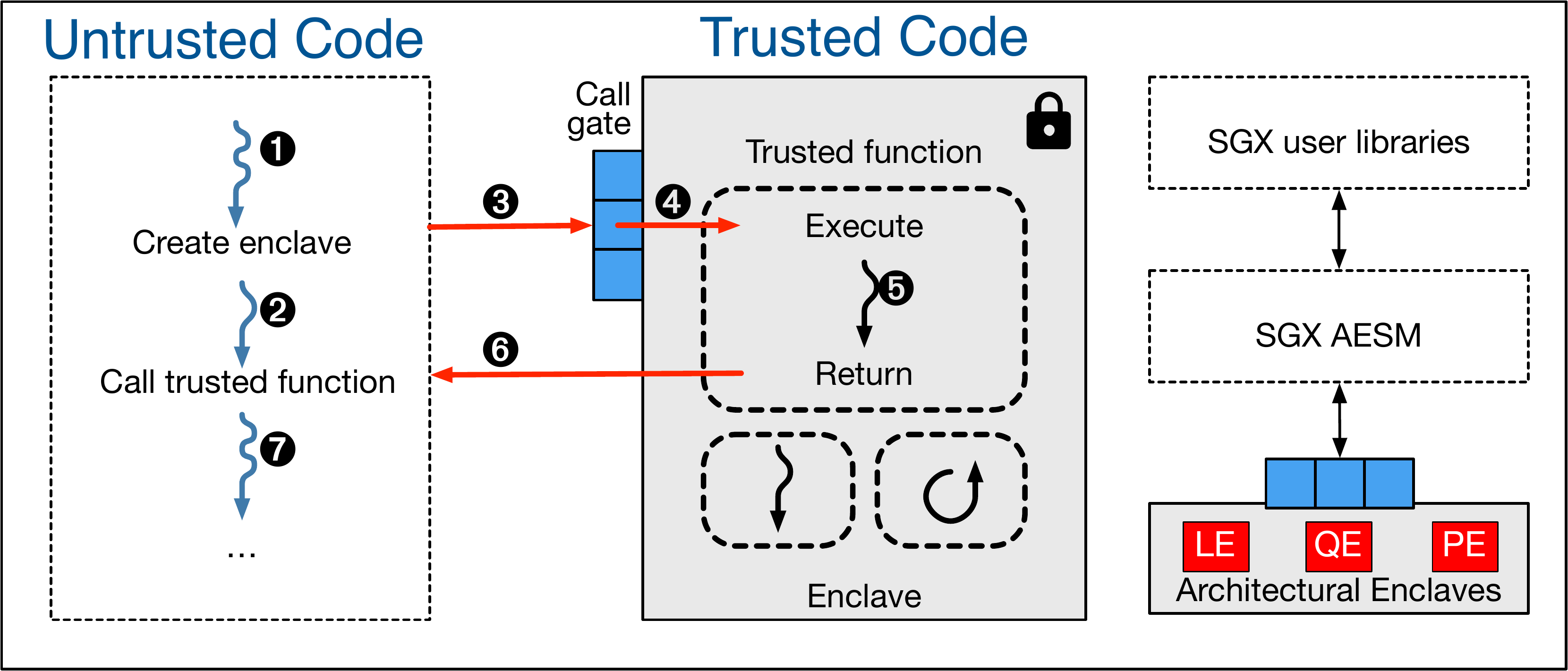}
  \caption{SGX core operating principles.}
  \label{fig:sgx}
\end{figure}

The \ac{SGX} mechanisms are depicted in \autoref{fig:sgx}.
They allow applications to access and process confidential data from inside the enclave.
The architecture guarantees that an attacker with physical access to a machine cannot tamper with the application data or code without being noticed.
Essentially, the \ac{CPU} package represents the security boundary.
Moreover, data belonging to an enclave is automatically encrypted, for 
confidentiality, and its digests inserted in a hash tree, for integrity and 
freshness.
A memory dump on a victim’s machine will only produce encrypted data.
A custom \emph{remote attestation protocol} allows to verify that a particular version of a specific enclave runs on a remote machine, using a genuine Intel processor with SGX enabled.
An application using enclaves must ship a signed (not encrypted) shared library (\ie a shared object file in Linux) that can possibly be inspected by malicious attackers.
Data stored in enclaves can be saved to persistent storage, protected by a seal key.
This allows to store sensitive data on disk, waiving the need for a new remote attestation every time the SGX application restarts.

There are three memory areas in which data relative to an \ac{SGX} enclave may be stored~\cite{costan2016intel}.
Data inside the \ac{CPU} package (registers, level 1-3 caches, etc.) is stored in plain-text, as long as the processor is in enclave mode and executes the enclave owning the data.
\ac{SGX} enclaves can also use a dedicated subset of system memory called the \ac{EPC}.
The \ac{EPC} is split into pages of \SI{4}{\kibi\byte} and exists in \ac{PRM}, a range of system memory that is inaccessible to other programs running on the same machine, including privileged software such as the operating system.
It is a small memory area; current hardware supports at most \SI{128}{\mebi\byte}.
This size is configurable via \ac{UEFI} parameters, but a reboot is required to apply the change.
Note also that the \ac{EPC} is shared among all the applications executing inside enclaves, hence being a very scarce and highly contended resource.
Future releases of \ac{SGX} will relax this size limitation~\cite{mckeen2016intel}.
Our evaluation shows, by means of simulation, how changes in the next version of \ac{SGX} would impact the performances of a scheduler, for example in terms of turnaround time for submitted jobs (see \autoref{subsec:eval:sgx}).

Only \SI{93.5}{\mebi\byte} out of \SI{128}{\mebi\byte} can effectively be used by applications (for a total of \num{23936} pages), while the rest is used for storing \ac{SGX} metadata.
In order to provide more flexibility, \ac{SGX} implements a paging mechanism.
It allows to page-out portions of trusted memory into regular system memory.
Any access to an enclave page that does not reside in the \ac{EPC} triggers a page fault.
The \ac{SGX} driver interacts with the CPU to choose the pages to evict.
The traffic between the CPU and the system memory is protected by the \ac{MEE}~\cite{gueron2016memory}, also in charge of providing tamper resistance and replay protection.
If a cache miss hits a protected region, the \ac{MEE} encrypts or decrypts data before sending to, respectively fetching from, the system memory and performs integrity checks.

The execution flow of a program using \ac{SGX} enclaves works as follows.
First, an enclave is created (see \autoref{fig:sgx}-\ding{202}) by the \emph{untrusted} part of the application.
It must then be initialized using a \emph{launch token}.
Intel's \ac{LE} can help in fetching such token.
Access to the \ac{LE} and other architectural enclaves, such as the \ac{QE} and the \ac{PE} is provided by the Intel \ac{AESM}.
SGX libraries provide an abstraction layer for communicating with the \ac{AESM}.
As soon as a program needs to execute a trusted function (see \autoref{fig:sgx}-\ding{203}), it executes the \texttt{ecall} SGX primitive (\ding{204}).
The call goes through the \ac{SGX} call gate to bring the executing thread inside the enclave (\ding{205}).
After the function is executed in the trusted environment (\ding{206}), it calls the return instruction (\ding{207}), before giving the control back to the caller (\ding{208}). 
 \section{Trust Model}
\label{sec:model}

We assume that our SGX-enabled orchestrator is deployed on the premises of a given cloud provider.
Providers show an honest-but-curious behavior.
They are interested in offering an efficient service to customers, mostly for selfish economic reasons (\eg providers want to maximize the number of executed jobs per unit of time, but they will not deliberately disrupt them).
However, providers do not trust their customers, especially not the resource usage declarations they specify at deployment time.
Nevertheless, providers trust their own infrastructure, namely the operating system running on the cluster nodes, the SGX driver, as well as Kubernetes itself.

Customers rely on the infrastructure offered by cloud providers, but do not trust it.
Notably, they trust Intel \ac{SGX} to prevent providers from inspecting the contents or tampering with jobs deployed on their infrastructure.
In turn, customers might try to allocate more resources than what they requested, either at the cost of providers or their tenants.

In previously existing cloud deployments, customers had to blindly trust providers not to pry into their jobs, nor tamper with them.
Conversely, providers had to trust that their customers do not overstep their share of \ac{EPC}, as they had no means to limit it.
Thanks to the novel work described in this paper, providers can now effectively supervise \ac{SGX}-related resources by enforcing \ac{EPC} allocation limits, thence rendering SGX cloud deployments practical.
As a consequence, customers can enjoy truthful job execution in the cloud without having to trust their provider.
 \section{Architecture}
\label{sec:architecture}

\begin{figure}[tpb]
	\centering
	\includegraphics[width=\linewidth,trim=0 0 0 0]{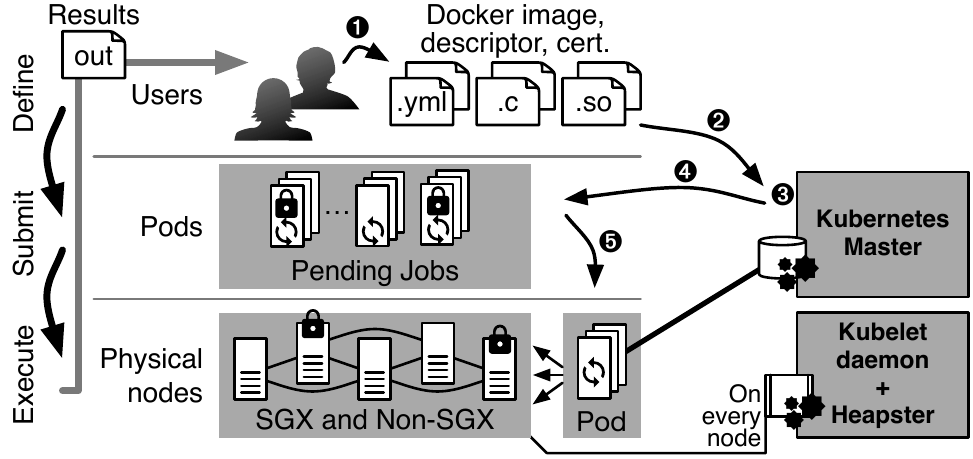}
	\caption{Architecture and workflow of the SGX-aware scheduler.}
	\label{fig:architecture}
\end{figure}

This section describes the architecture of our system, as depicted in \autoref{fig:architecture}.
The complete system implements a container orchestrator that can efficiently schedule SGX-enabled jobs, as well as regular jobs, on a heterogeneous cluster.
We assume that SGX-enabled jobs execute entirely in enclaves, minus a part responsible for bootstrapping the SGX enclaves.
These assumptions match those of state-of-the-art SGX systems, \eg SCONE~\cite{arnautov2016scone} or Graphene-SGX~\cite{tsai2017graphene}.

One important aspect of the system is its ability to guarantee that jobs submitted to a given host always fit within its \ac{EPC} memory limits.
This is of particular relevance to avoid major performance penalty~\cite{arnautov2016scone}.
To achieve this goal, we use a monitoring layer, implemented by means of daemons executing on each node of the cluster.
They accurately measure individual \ac{EPC} utilization of the jobs submitted to our system.
These measures are periodically sent to a central node, in charge of storing and analyzing them.
We rely on a time-series database for this task, as they have proved to be an efficient tool to implement queries over moving sliding windows~\cite{Pelkonen:2015:GFS:2824032.2824078,faloutsos1994fast,huang1999adaptive}, such as the ones we use in our scheduling policy.

Users submit their jobs by specifying the name of the container image (see \autoref{fig:architecture}-\ding{202}).
They need to indicate the amount of \ac{EPC} memory required by their jobs at this time.\footnote{As the allocation of \ac{EPC} memory has to be done at program initialization, this value is hard-coded in the SGX-enabled binary and could be extracted directly from it. We rely on the user's specification for convenience reasons.}
Apart from these parameters, clients rely on the regular \acp{API} of the orchestrator.
The image is initially pulled from a public or private container registry.
Afterwards, it is submitted into the scheduler's queue as a pending job (\ding{203}).

\textbf{Scheduling algorithm.} Our scheduling algorithm works as follows.
First, the container orchestrator fetches the list of pending jobs currently in the queue (\ding{204}).
It takes their memory allocation requests into account, both in terms of standard memory and \ac{EPC}.
At the same time, it fetches accurate, up-to-date metrics about memory usage across all nodes in the cluster.
This is done by executing a sliding-window query over the time-series database.
The scheduler then combines the two kinds of data to filter out job-node combinations that cannot be satisfied, either due to hardware compatibility (\ie SGX-enabled job on a non-SGX node), or if the job requests would saturate a node.
The next step depends on the concrete container placement policy configured for the cluster.
In particular, we enable support for SGX measures in two well-known placement policies.

When \emph{binpack} is in use, the scheduler always tries to fit as many jobs as possible on the same node.
As soon as its resources become insufficient, the scheduler advances to the next node in the pool.
The order of the nodes stays consistent by always sorting them in the same way.
In the case of a standard job, we sort SGX-enabled nodes at the end of this list, to preserve their resources for SGX-enabled jobs.

Conversely, the main goal of the \emph{spread} strategy is to even out the load across all nodes.
It works by choosing job-node combinations that yield the smallest standard deviation of load across the nodes.
Like \emph{binpack}, it only resorts to SGX-enabled nodes for non-SGX jobs when no other choice is possible to execute the job.
After the policy selection is made, the scheduler communicates the computed job-node assignments to the orchestrator (\ding{205}).
It then handles the actual deployment of jobs towards the various nodes (\ding{206}).

There may be jobs that cannot be fitted in the cluster at the time of their submission.
The orchestrator keeps a persistent queue of pending jobs (\ding{204}); the scheduler periodically checks for the possibility to schedule some of them, applying a \ac{FCFS} priority.

\section{Implementation Details}
\label{sec:implementation}
In this section we provide insights on the implementation of the components of our architecture.
Although several mainstream container orchestrators exist today, we decided to build our implementation on top of the open-source Kubernetes\footnote{We also report that our initial attempt was based on Docker Swarm. However, its lack of hooks to extend the architecture and poor developer documentation convinced us to look for a different solution.} container orchestrator~\cite{kubernetes-components}.
Likewise, the entire source code of our implementation is released as open-source software~\cite{github:sgx-scheduler}.
The components that we add to Kubernetes' architecture interact with it using its public \ac{API}.
This approach further facilitates the integration into future versions of Kubernetes.

In the remainder of this section, we provide implementation details for: our \emph{device plugin} (\autoref{subsec:device-plugin}), our custom SGX-aware scheduler (\autoref{subsec:custom-scheduler}), and how we implemented SGX metric probes (\autoref{subsec:metrics-probe}).
We then highlight how we enforce SGX-related resource usage limits (\autoref{subsec:kubelet}).
Moreover, we describe our extensions to the Intel SGX driver for Linux (\autoref{subsec:sgx-driver}).
Finally, we present our base Docker image that SGX application developers can use as a base for their applications (\autoref{subsec:docker-sgx-image}).

\subsection{Kubernetes \textit{device plugin}}
\label{subsec:device-plugin}
The first component is a \textit{device plugin} that allows to mark a Kubernetes node as able to execute SGX instructions. 
The plugin can be used directly by the Kubelet~\cite{kubernetes-components} node agent, since Kubernetes 1.8~\cite{device-plugins}.
Its original design was motivated by the developers' need to access \acp{GPU}. 
Its intent is to expose system devices (\ie, those available in the \texttt{/dev} pseudo-filesystem) directly within Kubernetes. 
Opportunely, applications implemented using the Intel \ac{SDK} for Linux are given access to SGX by means of a pseudo-file in \texttt{/dev}.
Communication between Kubelet and the \emph{device plugin} leverages \ac{gRPC}~\cite{grpc}.
Our \emph{device plugin} checks for the availability of the Intel SGX kernel module on each node and reports it to Kubelet.
Kubelet notifies the master node about the availability of an \enquote{SGX} resource on that node.

The philosophy behind \textit{device plugins} is to register one resource item (\eg one graphics processing card, one FPGA board, etc.) per physical device.
In the case of SGX applications, there is only one pseudo-file registered per processor.
However, SGX allows multiple enclaves to be executed at the same time, sharing the EPC.
Exposing only one resource item would have utterly limited the usefulness of our contribution, as only one SGX-enabled set of containers (or \emph{pod} in the Kubernetes terminology) could have been scheduled on a physical host at any given time.
We solve this problem by exposing each EPC page as a separate resource item.
By exposing the EPC as multiple independent \enquote{devices}, several pods can be deployed and share a single node, thus supporting the execution of several SGX applications at once.
Despite the great amount of resources created with this scheme, we did not notice any perceptible negative influence on performance.

Although SGX allows over-commitment of its protected memory via paging, doing so leads to severe performance drops up to $\num{1000}
\times$~\cite{arnautov2016scone}.
Therefore, we deliberately prevent over-commitment of the EPC, in order to preserve predictable performance for all pods deployed in the cluster.

Kubernetes will then proceed and mount the SGX device file inside each pod that requested at least one share of EPC.
Therefore, end-users must declare that their SGX-enabled pods use some amount of the \enquote{SGX} resource.
In the case of Kubernetes, this is done by filling in the resource requests and limits fields of the pod specification.
Resource requests are used by the scheduler to dispatch SGX-enabled pods towards a suitable node.
Limits are transmitted to our modified SGX driver for strict enforcement (see \autoref{subsec:kubelet}).

\subsection{SGX-Aware Scheduler}
\label{subsec:custom-scheduler}
Once nodes have been configured with our \emph{device plugin}, they are ready to accept pods using Intel SGX instructions.
However, Kubernetes' own scheduler only relies on values communicated in the resource requirements of each pod. 
Given the restrained capacity of the EPC, it is imperative to maximize its utilization factor.
To do so, the scheduler must consider the actual usage metrics of the cluster, collected at runtime.  
Our scheduler can decide on scheduling actions based on actual measured memory usage (for the EPC as well as regular memory).
Metrics regarding the regular memory are collected by Heapster~\cite{heapster}, while SGX-related metrics are gathered using our custom probes (see \autoref{subsec:metrics-probe}).

We implemented the previously described scheduling strategies (\emph{binpack} and \emph{spread}) in a non-preemptive manner, following the same scheme as Kubernetes' default scheduler.
The scheduler itself is packaged as a Kubernetes pod.
This allows it to execute with the same privileges as the default scheduler.
It also provides us with seamless migrations and crash monitoring features, as for any pod.

Kubernetes supports multiple schedulers to concurrently operate over the same cluster.
It is therefore possible to deploy our scheduler with both of its strategies, in parallel to the default, non SGX-aware one.
Comparative benchmarking is thus made easier, as each pod deployed to the cluster can specify which scheduler it requires.
We assume that, in production deployments, only one variant of our SGX-aware scheduler will be deployed as default scheduler to prevent conflicts between schedulers.

\subsection{SGX metrics probe}
\label{subsec:metrics-probe}

The proposed scheduling algorithm relies on metrics directly fetched from nodes of the cluster.
Kubernetes natively supports Heapster~\cite{heapster}, a lightweight monitoring framework for containers.
We configured Heapster to collect such metrics on each node and subsequently store them into an InfluxDB~\cite{influxdb} time-series database.
Then, the system performs InfluxQL queries~\cite{influxdb-ql} against the database.
\autoref{lst:influxql} shows how to report the total size of EPC memory used in the last 25 seconds per pod, then grouped by node and summed up.

\begin{listing}
\begin{minted}{SQL}
SELECT SUM(epc) AS epc FROM
  (SELECT MAX(value) AS epc FROM "sgx/epc"
    WHERE value <> 0 AND time >= now() - 25s
    GROUP BY pod_name, nodename
  )
GROUP BY nodename
\end{minted}
\caption{InfluxQL query.}
\label{lst:influxql}
\end{listing}

We have implemented an SGX-aware \textit{metrics probe} to gather EPC usage metrics from our modified Intel SGX driver (see \autoref{subsec:sgx-driver}).
These metrics are pushed into the same InfluxDB database used by Heapster.
This allows our scheduler to use equivalent queries for SGX- and non SGX-related metrics.

The probe is deployed on all SGX-enabled nodes using the \textit{DaemonSet} component~\cite{kub-deamonset}.
The distinction between standard and SGX-enabled cluster nodes is made by checking for the EPC size advertised to Kubernetes by the \emph{device plugin}.
Finally, we leverage Kubernetes itself to automatically handle the deployment of new probes when adding physical nodes to the cluster, as well as their management in case of crashes.

\subsection{Enforcing limits on EPC usage}
\label{subsec:kubelet}

SGX allows over-commitment of its primary cache, the EPC.
However, this feature comes with an important performance penalty for user applications.
It is imperative for a cloud provider that wants to guarantee a fair share of resources to make sure that multiple containers share the EPC in a respectful way.
In Kubernetes, the users specify the limits for each type of resource that their pods use.
These values are later used for several purposes, \eg, accounting and billing the reserved resources.
It is in the interest of the infrastructure provider to make sure that co-hosted containers do not contend on the same resource.
A user with malicious intents could advertise lower amounts of resources than what his pods actually use.
For this reason, it is crucial to enforce the limits advertised in the specification of each pod.
In our particular context, we focus on limits related to the EPC usage.

We implement proper limits enforcement by modifying two existing components of our architecture: (i) the SGX driver provided by Intel (see \autoref{subsec:sgx-driver}), and (ii) Kubelet, the daemon running on each node of a Kubernetes cluster.
The SGX driver will deny the initialization of any enclave that exceeds the share of pages advertised by its enclosing pod.

\textbf{Linux cgroups.} The proper way to implement resource limits in Linux is by adding a new \emph{cgroup} controller to the kernel~\cite{cgroups}.
This represents a substantial engineering and implementation effort, affecting several layers of our architecture.
Modifications would be required in Kubernetes, Docker (which Kubernetes uses as container runtime) and the Linux kernel itself.

We considered a simpler, more straightforward alternative.
Namely, we use the \emph{cgroup} path as a pod identifier.
The rationale behind the choice of this identifier is as follows:
(i)~it is readily available in Kubelet and in the kernel;
(ii)~all containers in a pod share the same \emph{cgroup} path, but distinct pods use different ones;
(iii)~the path is available before containers actually start, so this allows the driver to know the limits applicable to a particular enclave on its initialization.

In order to communicate limits from Kubernetes to the SGX driver, we added 16 lines of Go code and 22 lines of C code to Kubelet, using \emph{cgo}~\cite{cgo}.
These additions communicate a new \emph{cgroup path}--\emph{EPC pages limit} pair each time a pod is created.
Finally, the communication channel uses a new input/output queryable (\textit{ioctl})~\cite{man:ioctl} added to the SGX driver detailed in \autoref{subsec:sgx-driver}.

\subsection{Modified Intel SGX Driver}
\label{subsec:sgx-driver}

We modified the Linux kernel driver~\cite{github:driver} provided by Intel.
Our modifications revolve around two closely-related topics: gathering usage data to improve scheduling decisions, and enforcement of resource usage limits.
We offer access to the total number of EPC pages, as well as their current usage status by way of module parameters.
They are accessible using the usual Linux filesystem interface, below the \texttt{/sys/module/isgx/parameters} path.
Values can be retrieved through two pseudo-files: \texttt{sgx\_nr\_total\_epc\_pages} (total amount of pages on the system) and \texttt{sgx\_nr\_free\_pages} (amount of pages not allocated to a particular enclave).

Additionally, the EPC usage can be probed at a finer granularity, in a per-process manner. 
To do so, we created a new input/output queryable using the \texttt{ioctl} function~\cite{man:ioctl} available in Linux.
This control reports the number of occupied EPC pages given to a single process, described by its \textit{process identifier}.
This metric is helpful to identify processes that should be preempted and possibly migrated, a feature especially useful in scenarios of high contention.

We created a second \textit{ioctl} to communicate resource usage limits (see \autoref{subsec:kubelet}).
Each pod deployed in our cluster needs to advertise the number of EPC pages it plans to use.
This number is then shared between Kubelet and the driver by issuing an \textit{ioctl} at the time of pod creation.
The driver makes sure that limits can only be set once for each pod, therefore preventing the containers themselves from resetting them.
In the current version of SGX, enclaves must allocate all chunks of protected memory that they plan to use at initialization time.
We add a couple lines of code in \texttt{\_\_sgx\_encl\_init} to call a function that checks whether to allow or deny a given enclave initialization.
Internally, it compares the number of pages owned by the enclave to the limits advertised by its enclosing pod.

The implementation of these features consists in \num{115} lines of C code on top of the latest Intel SGX release for Linux~\cite{github:driver_modified}.

\subsection{Base Docker Image and Intel \ac{SDK} for SGX}
\label{subsec:docker-sgx-image}

When an application wants to use the processor features offered by Intel SGX, it has to operate in enclave mode.
Before it can switch to this mode, the program has to pre-allocate all the enclave memory that it plans to use.
The particular x86 instructions that can reserve enclave memory can only be executed by privileged code running in \textit{ring~0}~\cite{costan2016intel}.
Under the GNU/Linux operating system, only the kernel and its modules are allowed to call these instructions.
In a containerized context, the kernel and, by extension, its modules are shared across all containers.
In the case of running SGX-enabled applications in containers, this implies that it is required to set-up a communication channel between the container and the \textit{isgx} module.
In Docker, this can be done by mounting the \texttt{/dev/isgx} pseudo-file exposed by the host kernel directly into the container.
While it is possible to create SGX-enabled applications that directly interface with the kernel module, the Intel SGX \ac{SDK} provides an easier path to SGX application development.
Programs that are using this \ac{SDK} rely on the \ac{PSW}~\cite{sgx-install-guide}.
Hence, we created a Docker image that allows SGX-enabled applications developed using the official \ac{SDK} to be executed seamlessly within Docker containers.
The image is publicly available from Docker Hub~\cite{dockerhub:container}.
 \section{Evaluation}
\label{sec:evaluation}
This section reports on a detailed evaluation of our SGX-aware scheduler and its subcomponents.
First, we describe our experimental settings in \autoref{subsec:eval:settings}.
Then, in \autoref{subsec:eval:trace}, we characterize the Google Borg trace and the simplifications done to adapt it to the scale of our cluster.
We characterize the synthetic workload that we use in \autoref{subsec:workload-description}.
\autoref{subsec:eval:sgx} shows the performance penalties induced by SGX.
A comprehensive evaluation of the scheduler itself follows in \autoref{subsec:scheduler-evaluation}, with \autoref{subsec:limits-evaluation} concentrating on measuring the effectiveness of strictly enforcing resource usage limits.
We end our evaluation by investigating how our work will be affected by the release of \ac{SGX}~2 (\autoref{subsec:sgx2}).
 
\subsection{Evaluation settings}
\label{subsec:eval:settings}
Our cluster consists of 5 machines.
The first 3 machines are Dell PowerEdge R330 servers, each equipped with an Intel Xeon E3-1270\,v6 \ac{CPU} and \SI{64}{GiB} of RAM.
One of these machines acts as Kubernetes master, while the remaining are regular Kubernetes nodes.
The two remaining nodes are SGX-enabled machines, also acting as nodes in the Kubernetes cluster.
These machines feature an Intel i7-6700 CPU and \SI{8}{GiB} of RAM.
SGX is statically configured to reserve \SI{128}{MiB} of RAM for the EPC.
The machines are connected to a \SI{1}{\giga\bit\per\second} switched network.
We use Kubernetes (v1.8), installed on top of Ubuntu 16.04.
We enabled the Kubernetes \textit{device plugin} alpha feature on all the machines.

\subsection{The Google Borg Trace}
\label{subsec:eval:trace}

\begin{figure}[t]
	\centering
	\includegraphics[width=\figurewidth]{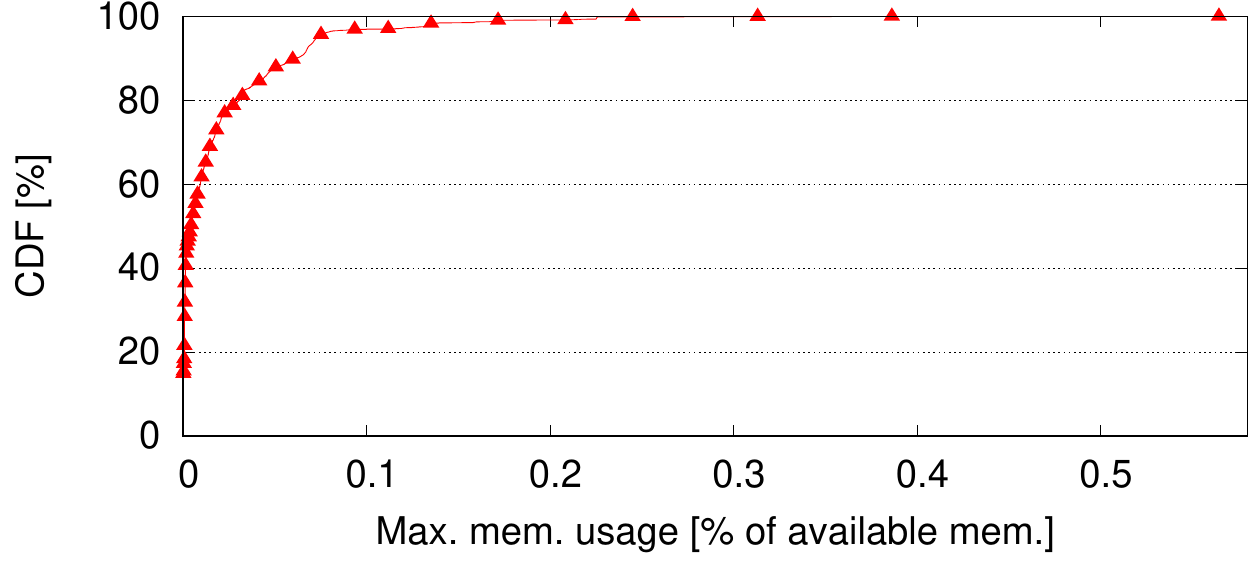}
	\caption{Google Borg trace: distribution of maximal memory usage.}
	\label{eval:borg:maxmem_usage}
\end{figure}

\begin{figure}[t]
	\centering
	\includegraphics[width=\figurewidth]{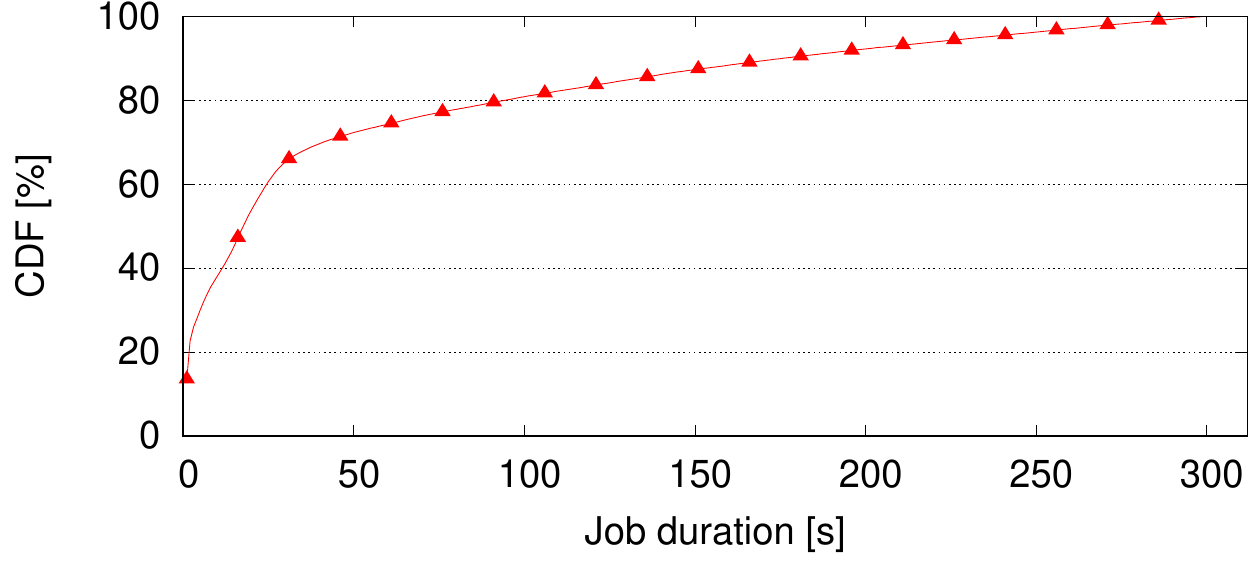}
	\caption{Google Borg trace: distribution of job duration.}
	\label{eval:borg:jobs_duration}
\end{figure}

\begin{figure}[t]
	\centering
	\includegraphics[width=\figurewidth]{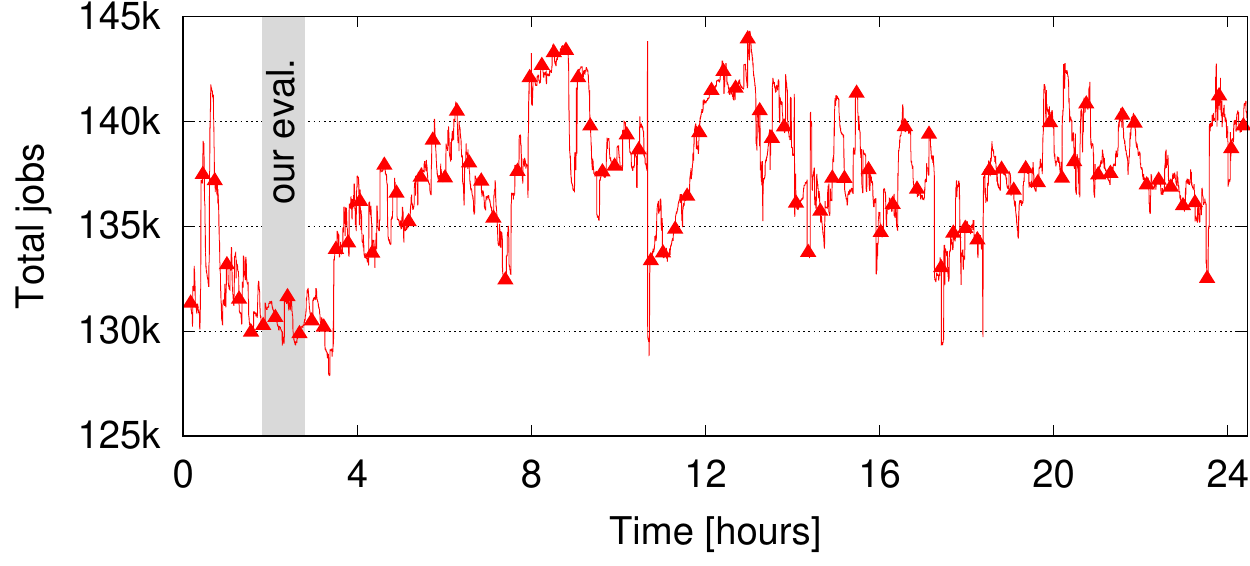}
	\caption{Google Borg trace: concurrently running jobs during the first \SI{24}{\hour}.}
	\label{eval:borg:running_jobs}
\end{figure}

Our evaluation uses the Google Borg Trace~\cite{clusterdata:Wilkes2011, clusterdata:Reiss2011}.
The trace was recorded in 2011 on a Google cluster of about \num{12500} machines.
The nature of the jobs in the trace is undisclosed. 
We are not aware of any publicly available trace that would contain SGX-enabled jobs.
Therefore, we arbitrarily designate a subset of trace jobs as SGX-enabled.
In the following experiments, we insert various percentages of SGX jobs in the system: from \SI{0}{\percent} of SGX jobs (only standard jobs), and then increasing by \SI{25}{\percent} steps, until \SI{100}{\percent} (only SGX jobs).

The trace reports several metrics measured for the Google jobs. 
We extract the following metrics out of it: \textit{submission time}, \textit{duration}, \textit{assigned memory} and \textit{maximal memory usage}.
The submission time is crucial to model the same arrival pattern of the jobs in our cluster.
The run time of each job matches exactly the one reported in the trace, set aside system clock skews. 
We use the assigned memory as the value advertised to Kubernetes when submitting the job to the system.
However, the job will allocate the amount given in the maximal memory usage field.
We believe this creates real-world-like behavior \textit{w.r.t.} the memory consumption advertised on creation compared to the memory that is actually used.

The trace specifies the memory usage of each job as a percentage of the largest memory capacity in Google's cluster (without actually reporting the absolute values).
In our experiments, we set the memory usage of SGX-enabled jobs by multiplying the memory usage factor obtained from the trace to the total usable size of the EPC (\SI{93.5}{MiB} in our case).
As for standard jobs, we compute their memory usage by multiplying them to \SI{32}{GiB}.
The rationale behind this choice is that it is the power-of-2 closest to the average of the total memory installed in our test machines.
Moreover, we think that it yields amounts that match real-world values.
\autoref{eval:borg:maxmem_usage} shows the amounts of memory allocations recorded in the trace.
Given the size of Google's cluster, we have to scale down the trace before being able to replay it on our own cluster setup.
We scale the trace down along two dimensions.

\textbf{Time reductions.} \autoref{eval:borg:jobs_duration} shows the \ac{CDF} of the duration of jobs found in the trace.
All jobs last at most \SI{300}{\second}.
Hence, \SI{1}{\hour} is sufficiently long to properly stabilize the system.
Instead of considering the full \num{29} days recorded in the trace, we use a 1-hour subset ranging from \SIrange{6480}{10080}{\second} extracted from the first 24 hours of execution (highlighted in light gray in \autoref{eval:borg:running_jobs}).
This slice of trace, while being the less job-intensive in terms of concurrent jobs for the considered time interval, still injects an intensive load on the cluster.
Hence, we also operate a frequency down-scaling of it, described next.

\textbf{Frequency reductions.} We sample every \num{1200}th job from the trace, to end up with a number of jobs big enough to cause contention in the system, but that does not clutter it with an incommensurable amount of jobs.
\autoref{eval:borg:running_jobs} displays the concurrent amount of running jobs recorded in the trace, before sampling.

\subsection{Matching trace jobs to deployable jobs}
\label{subsec:workload-description}

After processing the trace file, we get a timed sequence of jobs with their effective memory usage.
In order to materialize this information into actual memory or \ac{EPC} usage, our jobs are built around containers that run \textsc{Stress-SGX}~\cite{stress-sgx}, a fork of the popular \textsc{stress-ng}~\cite{stress-ng} stress tool.
Normal jobs use the original virtual memory stressor brought from \textsc{stress-ng}, while SGX-enabled jobs use the topical EPC stressor.
We specify parameters to allocate the right amount of memory for every job, in accordance with the values reported by the trace.

\subsection{Evaluation of SGX performance}
\label{subsec:eval:sgx}

The main sources of overhead for SGX enclaved executions are the transitions between protected and unprotected modes, and memory usage~\cite{arnautov2016scone,Pires:2016:SCBR,Brenner:2016:SecureKeeper,Orenbach:2017:Eleos}.
Additionally, startup time is longer than traditional executions, mainly due to support service initialization and memory allocation.
The Intel \ac{SDK}~\cite{sgx-install-guide} provides the \acf{PSW} that includes the \acf{AESM}.
As the name suggests, it is a service that eases the process of deploying enclaves, performing common tasks such as attestation and supporting the access to platform services, like obtaining trusted time and monotonic counters. 

By default, Docker---and more in general containers based on \textit{cgroups}---enforces strict limitations to what programs running inside the container can get access to.
It especially isolates the host from potentially malicious containers.
The isolation layer can be lifted by running containers in \textit{privileged} mode, a potentially dangerous and risky operation~\cite{bui2015analysis}, especially in a shared computing environment.
As we want to keep containers isolated, we need to have one instance of the \ac{PSW} running in each container.
Therefore, each SGX-enabled containerized process will suffer from a small initialization penalty, as it will need to wait for the \ac{PSW} to initialize before it can start its useful work.

An additional startup overhead is due to the enclave memory allocation, since all of it must be committed at enclave build time, to be measured for attestation purposes~\cite{mckeen2016intel}.

\begin{figure}[t]
	\centering
	\includegraphics[width=\figurewidth]{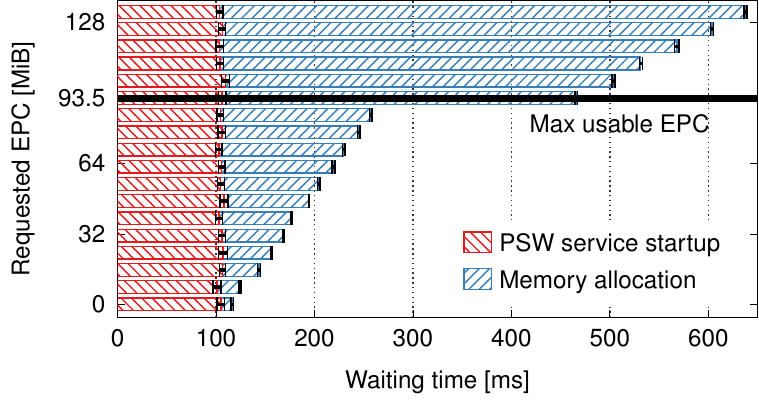}
	\caption{Startup time of SGX processes observed for varying EPC sizes.}
	\label{eval:sgx:startup}
\end{figure}

We start by quantifying the overhead for launching \ac{AESM} and allocating memory.
\autoref{eval:sgx:startup} shows the required average time required for \num{60} runs. 
Error bars represent the \SI{95}{\percent} confidence interval.
We omit measurements for standard jobs since they steadily took less than \SI{1}{\milli\second}, negligible compared to SGX ones.
As expected, the service startup time is virtually the same in all runs, accounting for about \SI{100}{\milli\second}.
On the other hand, memory allocation time shows two clear linear trends: before and after reaching the usable EPC memory limit. 
Until this limit, the time increase rate is \SI{1.6}{\milli\second\per\mebi\byte} after which it jumps to \SI{4.5}{\milli\second\per\mebi\byte}, plus a fixed delay of about \SI{200}{\milli\second}. 
Note that these times are just for \emph{allocating} memory, still without any real use.
Even higher overheads are expected when processes use it~\cite{arnautov2016scone}.
This experiment reinforces the importance of accurate SGX job scheduling to circumvent the soft-limits imposed by the EPC size.

\begin{figure}[t]
	\centering
	\includegraphics[width=\figurewidth]{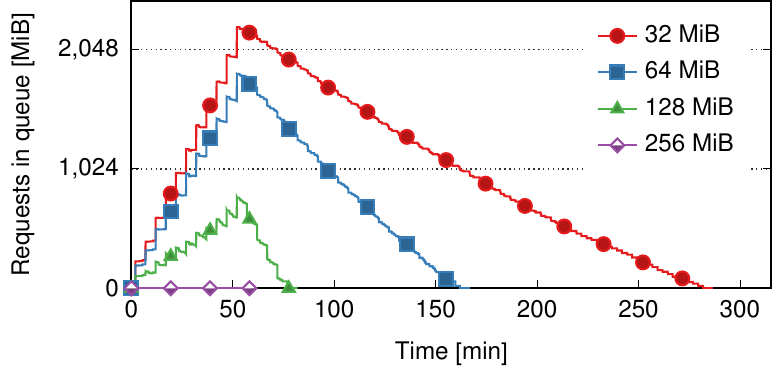}
	\caption{Time series of the total memory amount requested by pods in pending state for different simulated EPC sizes.}
	\label{eval:sgx:memory}
\end{figure}

Next, we evaluated the trace execution with different EPC sizes.
This particular run is based on simulation, but uses the exact same algorithms and behaves in the same way as our concrete scheduler.
It allows us to operate with various EPC sizes, including those that will be available with future SGX hardware.
\autoref{eval:sgx:memory} shows the amount of memory requested by pods in pending state along the time, with varying maximum reserved memory.
On the x-axis, we have the duration of the trace, while on the y-axis we see the total amount of EPC size requested by queued jobs that could not be immediately scheduled.
Looking at the extremes, we can notice the total absence of contention when the EPC accounts for \SI{256}{\mebi\byte}, finishing the batch execution in one hour, exactly as recorded in the trace.
Conversely, the trace takes \SI{4}{\hour}\,\SI{47}{\minute} when the maximum EPC memory usage is \SI{32}{\mebi\byte}.
In between, a \SI{64}{\mebi\byte} EPC would allow the trace execution to finish after  \SI{2}{\hour}\,\SI{47}{\minute}.
For \SI{128}{\mebi\byte}, the maximum EPC limit of current processors, the batch would conclude after \SI{1}{\hour}\,\SI{22}{\minute}.

This experiment, despite being dependent on this particular job trace, puts in evidence the benefits of having bigger protected memory sizes.
Having a processor capable of using \SI{256}{\mebi\byte} of EPC is theoretical at this stage.
Nonetheless, we expect SGX~2 (see \autoref{subsec:sgx2}) to allow larger EPC sizes, and consequently make SGX more appealing to cloud providers.

\subsection{Scheduler evaluation}
\label{subsec:scheduler-evaluation}

We conclude the evaluation of our SGX-aware scheduler by replaying the trace described in \autoref{subsec:eval:trace}.
All performance metrics are directly fetched from Kubernetes.

\begin{figure}[t]
	\centering
	\includegraphics[width=\figurewidth]{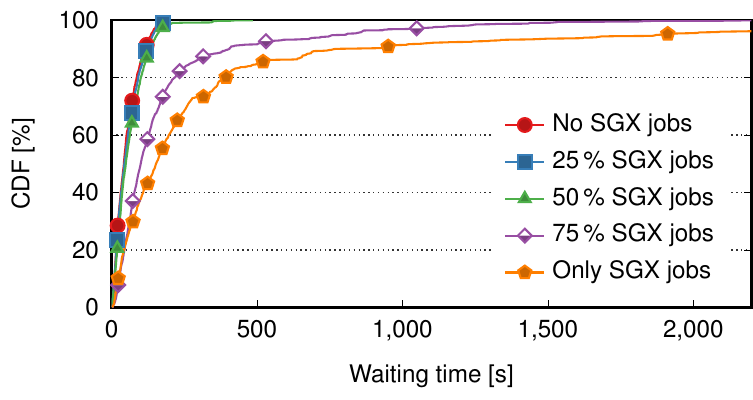}
	\caption{CDF of waiting times, using varying amounts of SGX-enabled jobs.}
	\label{eval:scheduler:waiting}
\end{figure}

\autoref{eval:scheduler:waiting} shows the \ac{CDF} of waiting times observed by jobs before their execution.
The waiting time refers to the period between the submission of the job to the orchestrator, and the instant when the job actually starts on a given node of the cluster.
In this experiment, we use the \emph{binpack} scheduling strategy.
As expected, the run that only uses standard memory (no SGX jobs) experiences relatively low waiting times.
On the other side, the pure SGX run waiting times go off the chart, due to much higher contention conditions.
The longest wait observed by a job is \SI{4696}{\second}, more than the total task duration given in the trace.
When \SIrange{25}{50}{\percent} of the jobs are SGX-enabled, waiting times are really close to the ones observed with a fully native job distribution.
This shows that incorporating a reasonable number of SGX jobs has close to zero impact on the scheduling.
Notably, we expect real-world deployments to include small percentages of jobs requiring SGX instructions, although this might change in the future.

\begin{figure}[t]
	\centering
	\includegraphics[width=\figurewidth]{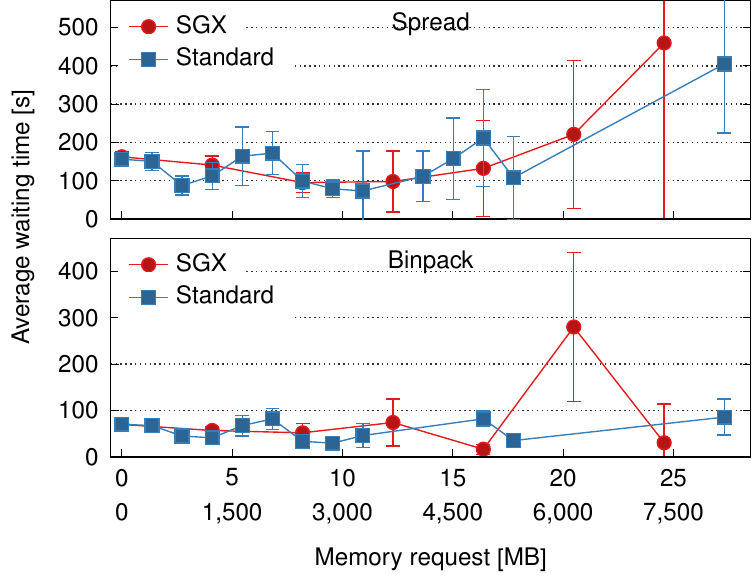}
	\caption{Waiting times for SGX and non-SGX jobs, using \emph{binpack} and \emph{spread} scheduling strategies, depending on the memory requested by pods.}
	\label{eval:scheduler:waiting_by_memory}
\end{figure}

\autoref{eval:scheduler:waiting_by_memory} depicts the waiting times observed in relation to the amount of memory requested by pods.
The top plot shows the results for the \emph{spread} strategy, while \emph{binpack} is shown at the bottom.
There are two rows of labels in the x-axis.
The top row with smaller values is applicable to SGX jobs while the bottom row is applicable to standard jobs.
All values are extracted from the same run with a \SI{50}{\percent} split between standard and SGX jobs.
The error bars are computed using the \SI{95}{\percent} confidence interval.
We can observe that the \emph{spread} strategy is consistently worse than \emph{binpack}.
Binpack also seems to handle bigger memory requests better.
SGX jobs show similar waiting times compared to standard jobs, save for one outlier in the \emph{binpack} plot.
This shows that our scheduler works well with both types of jobs.

\begin{figure}[t]
	\centering
	\includegraphics[width=\figurewidth]{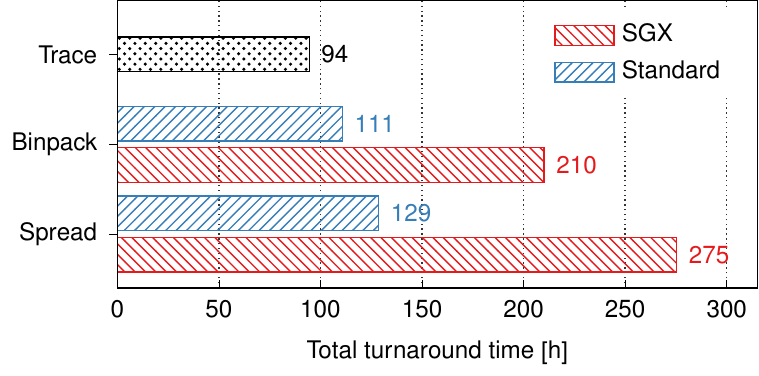}
	\caption{Sum of turnaround times for all jobs sent to the cluster, compared with the time reported by the trace.}
	\label{eval:scheduler:total_duration}
\end{figure}

Finally, \autoref{eval:scheduler:total_duration} shows the aggregated turnaround time for all jobs.
This metric refers to the duration elapsed between the instant a job was submitted to the moment when the job finishes and dies.
At the top of the figure, the dotted black bar (labeled \enquote{Trace}) represents the total useful job duration, as recorded in the trace.
The difference with the other results highlights the total waiting time for each of the different settings.
We use runs that only contain one type of job (either all SGX or regular jobs).
As we noted in \autoref{eval:scheduler:waiting_by_memory}, the \emph{binpack} strategy, in this specific setting and portion of the trace, achieves the best result (shorter turnaround time).
When using the \emph{binpack} strategy, SGX jobs need slightly less than twice the time of their non-SGX counterparts.
The total waiting time difference between the two kind of jobs is above this ratio, but the impact on the total turnaround time is limited to some extent.
Although a more in-depth evaluation of the trade-offs between the \emph{binpack} and \emph{spread} scheduling strategies would allow for a more comprehensive understanding of our setting, we believe that individual workload characteristics are the key factors when selecting a placement strategy.

Our decision to choose a multiplier of \SI{32}{\gibi\byte} for standard jobs and \SI{93.5}{\mebi\byte} for SGX-enabled jobs (see \autoref{subsec:eval:trace}) considerably affects the performance difference between standard and SGX jobs.
Indeed, our whole cluster has $2\times\SI{93.5}{\mebi\byte}=\SI{187}{\mebi\byte}$ of EPC memory (on the two SGX-enabled machines) compared to a total amount of $2\times\SI{64}{\gibi\byte}+2\times\SI{8}{\gibi\byte}=\SI{144}{\gibi\byte}$ of regular system memory.
This represents a difference of almost 3 orders of magnitude ($788\times$) between the two kinds of memory, whereas the difference between the scaling multipliers is only half of that ($350\times$).
Therefore, in relative terms, SGX jobs have close to $2\times$ less memory at their disposal compared to standard jobs.
As highlighted in \autoref{subsec:eval:sgx}, doubling the amount of memory drastically improves performance, which explains the performance gap.
We observe nonetheless that using a \SI{50}{\percent} split between standard and SGX jobs yields acceptable performance for both kinds of jobs.

\subsection{Impact of resource usage limits}
\label{subsec:limits-evaluation}

\begin{figure}[t]
	\centering
	\includegraphics[width=\figurewidth]{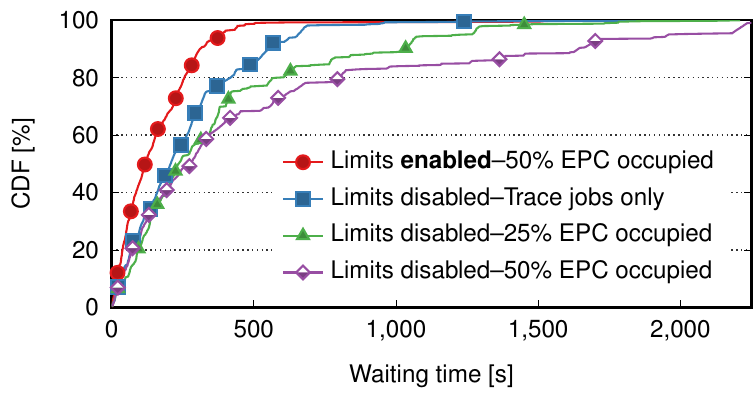}
	\caption{Observed waiting times when malicious containers are deployed in the system, with and without usage limits being enforced.}
	\label{eval:limits:cdf}
\end{figure}

One of the feature of our system is the strict enforcement of limits regarding per-container EPC memory consumption (see \autoref{subsec:kubelet}).
Resource usage limits are declared by the users themselves, and therefore could be inaccurate with regard to the real usage made by their containers.
We identify incentives to lead users into truthfully declaring resource usages: if the user declares too high a limit for his container, then the infrastructure provider will charge him for the additional resources.
On the other hand, declaring too low resource usages will lead to the container being denied service due to the enforcement of limits.
To show the potential damages that users could cause without this mechanism, we ran the same experiment as in \autoref{subsec:eval:trace}, but we add so-called malicious containers to the system.
We deploy as many of them as there are SGX-enabled nodes in the cluster.
The \textit{modus operandi} of these containers is to declare 1 page of EPC as limit and request in their pod specification, but actually use way more: up to \SI{50}{\percent} of the total EPC available on the machine they execute on.

The results, as presented in \autoref{eval:limits:cdf}, show that, without strict limits being enforced, most honest containers in the system suffer from longer waiting times.
Obviously, as the size of the allocations made by malicious containers increases, the effects suffered by honest containers grow as well.
Fortunately, we clearly see that enforcing limits on memory allocations annihilates the efforts of the malicious containers.
The reason why the run with malicious containers and limits enabled is better than the one that is just following the trace is because some jobs in the Google Borg trace actually try to allocate more memory than they advertise (\num{44} jobs out of \num{663} show this behavior).
When we strictly enforce memory limits, these jobs are immediately killed after launch.

\subsection{Compatibility with SGX~2}
\label{subsec:sgx2}

The hardware in current Intel processors only supports version 1 of SGX.
Intel has already published several design documents regarding the second version, SGX~2~\cite{mckeen2016intel, intel-architecture-manual}.
The \ac{SDK} and the driver adapted to SGX~2 have been recently published~\cite{sgx-install-guide}.
The most important feature that this new version introduces is dynamic EPC memory allocation.
Enclaves can ask the operating system for the allocation of new memory pages, and may also release pages they own.
Contrary to the current version, these operations can also be done during their execution.
Considering the limited amount of \ac{EPC} that is shared by every enclave running on the same node, this new feature can really improve resource utilization on shared infrastructures.

As far as our scheduler is concerned, we believe that only minor changes need to be performed to fully take these new possibilities into account.
Provided that Kubernetes nodes are deployed on SGX~2-compatible hardware, we think that our solution will work out-of-the-box.
The scheduler already uses up-to-date measurements regarding \ac{EPC} usage to come up with scheduling decisions.
Even when using SGX~1, variations of \ac{EPC} usage can already happen if a container launches multiple enclaves whose life-cycles are not harmonized.
The only part of our system that we have identified as not yet SGX~2-ready is our implementation of resource usage limits in the Intel \ac{SGX} driver.
We believe that the effort required to port it to the new revision of \ac{SGX} is modest.
 
\section{Related Work}
\label{sec:related}
The problem of scheduling jobs over a cluster of heterogeneous machines has always attracted a lot of research.

To protect sensitive hypervisor scheduling decisions, \textit{Scotch}~\cite{scotch:raid17} conveys information gathered in \ac{SMM} to enclaves. 
The authors implement a prototype on top of the Xen hypervisor, adding about \SI{14}{\percent} of overhead for each context switch and interrupt. Validation is done by checking the accuracy of accounting under resource interference-~\cite{Varadarajan:2012:RAI:2382196.2382228} and escape-~\cite{cloudburst:2009} controlled attacks. 
Although they provide results within \SI{2}{\percent} of the ground truth in such scenarios, there is no guarantee that measurements coming from \ac{SMM} are not tampered with during control switch to enclave entry points. 
Their focus is on the protection of probing data and preventing improper resource usage.
Instead, we deal with system support for scheduling SGX jobs based on their main contentious resource: EPC memory pages.

Similar to our work, \textit{ConVGPU}~\cite{Kang2017:ConVGPU} provides solutions for scheduling jobs based on memory contention in container-based virtualized environments.
Specifically, they provide a mechanism that shares the \ac{GPU} memory among multiple containers.
Just as the EPC, \ac{GPU} memory is limited. 
However, it is not possible to swap out memory once it is full, an event that usually leads to more severe issues than just performance degradation.
To avoid that, ConVGPU intercepts and keeps track of memory allocation calls to the CUDA \ac{API} by providing an alternative shared library to applications running within containers.
Whenever a request cannot be granted, it holds the application's execution by postponing the call return until there are available resources.
They evaluate the system using four strategies for the selection of which waiting application should be served first, and show low overall application running time overheads.
Essentially, they act reactively to potential memory contention issues, after container deployment, whereas we take scheduling decisions before, based on self-declared memory needs, and after deployment, based on probed data. 
Besides, they only take into account intra-node resource management, and leave distributed processing by integration with Docker Swarm for future work.

Checkpointing and migration of running processes closely relates to scheduling strategies. 
In this direction, Gu~\etal~\cite{Gu:2017:SGXMigration} tackle these issues for SGX enclaves.
The challenge lies in securely creating, transmitting and restoring an enclave checkpoint while preserving all security guarantees, while not introducing new attack vectors.
Checkpointing a running SGX process, however, already imposes some obstacles, since part of the enclave metadata is not even accessible by the enclave itself.
Moreover, for consistency reasons, one must ensure that all application threads do not continuously mutate the state of the job being migrated.
The authors deal with the first issue by \emph{inferring} the value of such unreadable metadata.
Then, they issue replay operations that lead to an identical state.
However, their approach rely on the cooperation of the untrusted operating system, and therefore check afterwards if it has behaved accordingly.
The problem of achieving a \emph{quiescent} point, when all threads are guaranteed to not modify the process state, is done by synchronization variables kept inside the enclave, and by intercepting its entry and exit points. 
By doing this, they force all threads to reach either a dormant or a spinning state
that will only be undone after restoring the enclave at the target node.
After successfully creating the checkpoint, they still have to provide means to 
ensure that it cannot be restored more than once (fork attack) 
nor that an old one can possibly be recovered (rollback attack).
That is solved by means of a migration key transmitted through secure channels built by leveraging SGX attestation and by a self-destroy approach, which prevents the enclave from being resumed after it was checkpointed.
Overall, authors~\cite{Gu:2017:SGXMigration} show a negligible performance overhead.
Such mechanism could eventually be integrated into our system, towards a 
globally optimized \ac{EPC} utilization through the migration of enclaves.
However, we stress that the problem of SGX enclave migration (online or offline) is considered orthogonal to ours.

Similar to our work, \emph{MixHeter}~\cite{zhang:2018:mixheter} also deals with scheduling in heterogeneous environments.
Since different sorts of applications benefit from distinct hardware capabilities (\eg \acp{GPU} for graph computing, RAM for sorting jobs, etc.), distributed system schedulers that deal with mixed workloads and take decisions disregarding this aspect may face poor performance.
To that end, \emph{MixHeter} proposes a scheduler based on \emph{or-constraints}.
The different resource requests are translated into algebraic expressions to be satisfied. 
If preferred resources are busy, non-preferred, but still compatible ones are used instead, which can 
maximize the overall performance.
They evaluate the system on a popular scheduler for distributed systems
processing frameworks and show performance improvements up to \SI{60}{\percent}.
We only consider SGX and non-SGX jobs as characterizing features for orchestration decisions. 
Our system would therefore benefit from such a scheme considering a broader 
range of hardware capabilities, assuming applications would support alternative solutions (\eg AMD \ac{SEV}~\cite{amd-sev}, \acp{TPM}~\cite{tpm}, ARM TrustZone~\cite{arm2009security}) in the absence of Intel \ac{SGX}.

\acresetall

\section{Conclusion}
\label{sec:conclusion}

In this paper we have proposed a novel orchestrator for containers running on heterogeneous clusters of servers, with and without Intel \ac{SGX} support.
This technology allows users to deploy their software data in the cloud without having to trust the providers.
In \ac{SGX} enclaves, software runs at almost native speed, unlike with cryptographic mechanisms that have severe limitations in terms of performance and features.

The challenge in such deployments is to schedule containers with security requirements to \ac{SGX} machines in priority, which are scarce, while at the same time carefully monitoring their usage of \ac{SGX} resources.
In particular, when exceeding the limited memory capacity of \ac{SGX} enclaves, performance starts degrading significantly so it is therefore important not to overload \ac{SGX} machines with too many resource-demanding containers.
To ensure proper monitoring of low-level \ac{SGX} metrics, we extended the \ac{SGX} Linux driver to gather statistics about the \ac{SGX} runtime and feed them into the orchestrator, based on Kubernetes.
We developed a complete prototype that we openly release~\cite{github:sgx-scheduler}, deployed it in a private cluster, and conducted a detailed evaluation using Google Borg traces.
Our findings reveal that the scheduler must carefully take the \ac{EPC} size into account to reduce the overall turnaround time.
Also, we observed that when half of the jobs in the workload are \ac{SGX}-enabled, there is virtually no impact on general performance.
Finally, our experiments show that there is a small bootstrap time that \ac{SGX} containers must be ready to tolerate at startup.

As part of future research directions, we plan to extend our orchestrator by integrating support for enclave migration as well as hybrid processes running trusted and untrusted code.
 
\section*{Acknowledgment}
The research leading to these results has received funding from the European Union’s Horizon 2020 research and innovation programme and was supported by the Swiss State Secretariat for Education, Research and Innovation (SERI) under grant agreement No 690111.

{
\printbibliography
}

\end{document}